\documentclass[preprint,showpacs,preprintnumbers,amsmath,amssymb]{revtex4}
\usepackage{graphicx}
\usepackage{dcolumn}
\usepackage{bm}
\usepackage{epsfig}
\usepackage{subfigure}

\newcommand{\bi}{\begin{itemize}}
\newcommand{\ei}{\end{itemize}}
\newcommand{\be}{\begin{eqnarray}}
\newcommand{\ee}{\end{eqnarray}}
\newcommand{\bbmatrix}{\left( \begin{array}}
\newcommand{\eematrix}{\end{array} \right)}

\begin{document}

\title{The metal insulator transition in cluster dynamical mean field theory: intersite correlation,
cluster size, interaction strength, and the location of the transition line}
\author{Chungwei~Lin and Andrew.~J.~Millis}
\affiliation{ Department of Physics, Columbia University \\
538W 120th St NY, NY 10027}

\begin{abstract}
To gain insight into the physics of the metal insulator transition and the
effectiveness of cluster dynamical mean field theory (DMFT) we have used
one, two and four site dynamical mean field theory  to   solve
a polaron model of electrons coupled to a classical phonon field. 
The cluster size dependence of the metal to polaronic insulator
phase boundary is determined along with electron spectral functions and cluster correlation functions.
Pronounced cluster size effects start to   occur in   the intermediate coupling  
region in which the cluster calculation leads to a gap and the single-site approximation does not.
Differences (in particular a sharper band edge) persist in the strong coupling regime.
A partial density of states
is defined encoding a generalized nesting property of the band structure; variations in this
density of states account for differences between the dynamical cluster approximation and the cellular-DMFT
implementations of cluster DMFT, and for differences in behavior between the single band 
models appropriate for cuprates and the multiband models appropriate for manganites. 
A pole or strong resonance in the self energy is associated with insulating states; 
the momentum dependence of the pole is found to distinguish between Slater-like and Mott-like mechanisms
for metal insulator transition.
 Implications for the theoretical treatment of doped manganites are discussed.
\end{abstract}

\maketitle

\section{Introduction}

Correlation-driven metal-insulator transitions are one of the fundamental issues in electronic
condensed matter physics \cite{Imada_98}. It has known for many years that in models of electrons moving
on a periodic lattice and subject to short-ranged interactions, insulating phases may occur if the number of electrons
per unit cell, $N$, is an integer or rational fraction. If $N$ is an even integer, then the Pauli principle makes 
insulating behavior possible even in the absence of electron-electron interactions (this is the case
of filled bands). If $N$ is not an even integer, interactions are required to obtain an insulator. 
Interactions may lead to a spontaneous breaking of translational symmetry, implying a unit cell larger 
by a factor of $\lambda$ than the crystallographic cell. If $\lambda$ is such that $\lambda N$ is an even
integer, insulating behavior is clearly possible; this is referred to as the "Slater mechanism" \cite{Slater_51}.
On the other hand insulating behavior may occur for $N$ not an even
integer, even in the absence of spontaneous translational symmetry breaking. This is often termed the
"Mott mechanism" \cite{Mott_37}.

Important new insights into metal-insulator transition physics have come with with the development of
dynamical mean field theory (DMFT) \cite{DMFT_96}. Single-site dynamical mean field theory, which
entirely neglects spatial correlations in the electron gas, is found to predict a zero-temperature
metal insulator transition as the local interaction $U$ exceeds a critical strength $U^{single-site}_c$ at density
$N=1$ \cite{DMFT_96}. More recently, cluster dynamical mean field theories \cite{Maier_05,Gull_08,Park_08}
which include some degree of intersite correlations have been shown to produce metal insulator
transitions at much smaller values of $U$ ($\sim 0.4 U^{single-site}_c$ when density $N=1$). However, the
physics of the metal-insulator transition observed in the cluster-DMFT methods remains imperfectly
understood, in part, because of the substantial computational expense of studying fully quantum cluster models. 

In this paper we present results of a systematic cluster DMFT study of a computationally simple 
polaron model, involving electrons coupled by an on-site density coupling to a dispersionless classical
oscillator. The model exhibits spectra and metal-insulator phase diagrams similar to those of
the interacting electron models of primary physical interest, but allows an 
extensive analysis. We are able to determine the relation between short ranged
correlations and the metal insulator phase boundary; elucidate the role of poles in the self energy,
determine the minimum cluster size needed to see the effects of short ranged interaction and clarify the origin of the differences
between dynamical cluster approximation (DCA) and cellular-DMFT approaches. 

The rest of this paper is organized as follows. In section II we present the formalism we will use.
We introduce the quantity "partial density of states" (PDOS) which is very important
in determining the importance of short ranged effects. In section III we review the polaron model, give some 
calculational details related to the impurity cluster problem, and discuss
different theoretical ways to measure the strength of short-ranged correlation.
In section IV we compare the results of one, two, and four-site cluster approaches
applied to the half-filled model. In particular we show that the momentum dependent
poles in self energies is the key to reduce the critical gap-opening $U$. 
Section V concerns how the poles in self energies depend on the temperature and interaction strength.
In section VI
we discuss doping dependence and give a criterion on the minimal cluster size required to capture the
short-ranged charge ordering effects. In section VII we compared the results from DCA and cellular-DMFT, and from
manganite and cuprate bands to illustrate the role of the partial density of states in cluster-DMFT calculations.
Finally in a conclusion we summarize our results and discuss their implications. Two appendices give
details of partial density of states and strong coupling calculations.

\section{General Formalism of cluster DMFT}
\subsection{Overview}
In this section we review the general formalism for cluster 
dynamical mean field theory and give the equations for two frequently used approaches -- cellular DMFT (CDMFT) and
dynamical cluster approximation (DCA), and define a "partial density of states" (PDOS) used in subsequent analysis.
Special attention is paid to the 2-site and 4-site DCA formalism on the square and cubic lattice.
The main focus of this section is on the self-consistency condition. 

\subsection{Cluster dynamical mean field formalism}
In an $N_s$-site cluster DMFT calculation, one solves a quantum impurity model which may be described as an 
$N_s$-site cluster coupled to a non-interacting electron bath. The solution is expressed in terms of the 
impurity Green's function and self energy - $\hat{G}^{imp}$, $\hat{\Sigma}^{imp}$ which are $N_s \times N_s$ matrices
in the space of cluster labels.
The dynamical mean field step requires relating components of $\hat{G}^{imp}$ and $\hat{\Sigma}^{imp}$ 
to some combinations of lattice Green's functions and self energies. 
There are two frequently used algorithms: the cellular-DMFT approach \cite{Kotliar_01}
which is defined in real space and the DCA approach \cite{Hettler_98} defined in momentum space. 

In the cellular-DMFT to a given lattice, one  chooses an $N_s$-site supercell which periodically
tiles the lattice and writes the Green's function as a matrix in this supercell, with entries
depending on momentum $\vec{k}$ in the reduced Brillouin zone (RBZ) of the supercell lattice.
One identifies sites of the quantum impurity model with sites of the supercell and take interactions
of the quantum impurity model to be those of the original model, but confined to a supercell. 
The self consistency equation is
\be
\hat{G}^{imp} = \int_{\vec{k}\in RBZ} (dk) \left[ \omega+\mu-\hat{t}_{\vec{k}} -\hat{\Sigma}^{imp} \right]^{-1}
\label{eqn:CDMFT_SC}
\ee
with $(dk) = (d^dk)/(2 \pi)^d$, $RBZ$ standing for the reduced Brillouin zone, and $\hat{t}_{\vec{k}}$
the $N_s \times N_s$ hopping matrix expressed in the basis of enlarged cell. In this approach 
the self-consistency equation is
$\hat{G}^{imp}_{ij} = G^{lat}(\vec{R}_i-\vec{R}_j)$ with $\{\vec{R}\}$ basis vectors of the enlarged cell.
The cellular DMFT method breaks the lattice translational symmetry
because the intersite correlations (expressed in $\hat{\Sigma}^{imp}_{ij}$)
within a supercell are $different$ from those between supercells. After the cellular-DMFT equations are solved
one may restore the translation invariance by periodizing the self energy. Different methods
have been proposed \cite{Stanescu_06}; the issue is however not relevant to the considerations of this paper.

The DCA approximation is constructed in momentum space. One tiles the Brillouin zone into $N_s$ equal area 
non-overlapping tiles which we label by the average momentum $\vec{K}_i$. One approximates the self energy
$\Sigma(\vec{k},\omega)$ by $\Sigma_{\vec{K}_i}(\omega)$ if $\vec{k}$ is in the tile corresponding to $\vec{K}_i$.
The $impurity$ cluster model is defined by states labelled by the discrete set of labels $\vec{K}_i$;
the impurity model interaction is taken to be the momentum-space lattice interaction
$U_{\vec{k}_1 \vec{k}_2 \vec{k}_3 \vec{k}_4}$ but with wavevectors restricted to the set $\vec{K}_i$.
In the $\vec{K}_i$-basis the impurity model is diagonal and the self consistency equation is 
\be
G^{imp}_{{\vec{K}_i}}(\omega) = N_s \times\int_{\vec{p} \in i} (dp) 
\frac{1}{\omega+\mu-\epsilon_{\vec{p}} - \Sigma_{{\vec{K}_i}} (\omega)}
\label{eqn:SC_DCA}
\ee
Similar to single-site DMFT, it is useful to define the partial density of states
(PDOS) for each $\vec{K}_i$ sector as 
\be
D_i(\epsilon) = N_s \times \int_{\vec{p} \in \vec{K}_i} (dp) \,\,\delta(\epsilon-\epsilon_{\vec{p}})
\label{eqn:Def_PDOS}
\ee
(note that the prefactor $N_s$ leads to $\int d\epsilon D_i(\epsilon) = 1$). Once the PDOS is given,
the self consistency equation is
\be 
G^{imp}_{\vec{K}_i}(\omega) = \int d\epsilon \frac{D_i(\epsilon)}
{\omega+\mu-\epsilon - \Sigma^{imp}_{{\vec{K}_i}}(\omega)}
\ee
In all cases we have studied, the $\vec{K}_i$ basis diagonalizes the cellular DMFT impurity model
($G$ and $\Sigma$) also. Therefore the cellular DMFT self consistency condition can be expressed in the form
of Eq(\ref{eqn:SC_DCA}) but the PDOS $D_i(\epsilon)$ will be different.
An example is given in Appendix A.

One notices that if all partial density of states are identical to the total density of states, 
the solution in single site DMFT $is$ a solution in DCA, i.e. 
$\Sigma^{imp}_{\vec{K}_i}$ is identical for all $\vec{K}_i$ ($\Sigma$ is purely local). 
A natural expectation is therefore that the separation between different sectors of PDOS
indicates the degree of the short-ranged correlation. In the subsequent sections we show that this is indeed the case.

While the DCA is naturally formulated in the cluster momentum basis $\vec{K}_i$, 
it is often convenient to define a real-space basis by introducing a set of vectors 
$\{ \vec{R} \}_{i=1 \sim N_s}$ satisfying
\be
\delta_{i,j} = \frac{1}{N_s} \sum_{l=1}^{N_s} e^{i (\vec{K}_i-\vec{K}_j)\cdot \vec{R}_l}
\label{eqn:determine_Ri}
\ee
(in this basis the on-site interaction term in the impurity model is identical to that in the lattice)
Note that the $\{\vec{R}_i\}$  are only determined up to a uniform translation 
$\{\vec{R}_i\} \rightarrow \{\vec{R}_i+\vec{A}\}$ with $\vec{A}$ a lattice vector.
Eq(\ref{eqn:determine_Ri}) enables one to get real-space components from the impurity problem
by the discrete Fourier transform
\be
\Sigma^{imp}_{{\vec{R}_m}}(\omega) &=& 
\frac{1}{N_s}\sum_{i=1}^{N_s} e^{-i \vec{K}_i \cdot \vec{R}_m} \Sigma^{imp}(\vec{K}_i, \omega)
\nonumber \\
G^{imp}_{{\vec{R}_m}}(\omega) &=& 
\frac{1}{N_s} \sum_{i=1}^{N_s} e^{-i \vec{K}_i \cdot \vec{R}_m} G^{imp}(\vec{K}_i, \omega)
\ee
where $\Sigma^{imp}(\vec{R}_m) = \Sigma^{imp}_{i,i+m}$ and $G^{imp}(\vec{R}_m) = G^{imp}_{i,i+m}$.
For $i$th momentum sector, the self energy is $\Sigma^{imp}_{{\vec{K}_i}}(\omega)$.

\begin{figure}[htbp]
\begin{center}
   \epsfig{file = 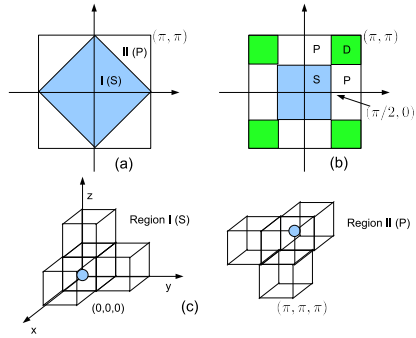,  width = 0.4\textwidth}
   \caption{(Color online) Partition of the Brillouin zone. (a) 2-site DCA on square lattice.  
   (b) 4-site DCA on square lattice.  (c) 2-site DCA on cubic lattice.
	}
   \label{fig:k_points}
\end{center}
\end{figure}

In cellular DMFT, different choices of $N_s$-site supercell are possible. In DCA different tilings of the
Brillouin zone are possible. A comparative survey of the different choices has not been undertaken
and would be worthwhile. It is natural to expect that tilings should respect the point symmetry
of the lattice as much as possible, i.e. if the lattice remains invariant under the transform 
$\vec{r} \rightarrow \vec{r}' =U\vec{r}$, 
the ideal tiling of the Brillouin zone obeys $\Sigma(\vec{k}) = \Sigma(U^{-1}\vec{k})$.
 In this paper we consider 2-site and 4-site tiling on
square and cubic lattices plotted in Fig(\ref{fig:k_points}).

{\it Notations}: Since there will be many subscripts floating around in this paper, we
explain the logic of our notation. For each quantity, like Green's function $G$, self energy $\Sigma$,
and Weiss function $a$, there are momentum and real-space components labeled by some subscript. 
In this paper, the real-space component is labeled by a number (0:on-site, 1:nearest neighbor..)
while the momentum-space sectors labelled by capital letters (S, P, D..).

\subsection{Formalism for 2-site DCA}
Now we apply the machinery just described to specific cases, first to the 2-site DCA in
the square lattice. After solving the 2-site impurity problem, one gets
(in the real space basis)
\begin{eqnarray}
\hat{G}^{imp} = \left( \begin{array}{cc}
    G_0 & G_1   \\    G_1 & G_0
\end{array}  \right)\,\,\,
\hat{\Sigma}^{imp} = \left( \begin{array}{cc}
    \Sigma_0 & \Sigma_1   \\    \Sigma_1 & \Sigma_0
\end{array}  \right)
 \label{eqn:G_2}
\end{eqnarray}

The partitioning of Brillouin zone in this case is given in Fig(\ref{fig:k_points}), so
two $\vec{K}$ points according to this division is $\vec{K}_I=0$, $\vec{K}_{II} = (\pi,\pi)$. 
We label region $I$ and $II$ or $S$ and $P$ sectors. Corresponding to $\vec{K}_I$ and 
$\vec{K}_{II}$, one gets $\vec{R}_0=0$ and $\vec{R}_1=(\pm1,0)$ or $(0,\pm1)$ from the requirement Eq(\ref{eqn:determine_Ri}).
The lattice self energy is related to $\hat{\Sigma}^{imp}$ by
\begin{eqnarray}
\Sigma_{DCA}(\vec{k}, \omega) = 
\left\{ \begin{array}{ll} 
\Sigma^{imp}_S = \Sigma_0 + \Sigma_1  & \mbox{for}\, \vec{k} \in \mbox{Region }I (S)\\ 
\Sigma^{imp}_P =\Sigma_0 - \Sigma_1   & \mbox{for}\, \vec{k} \in \mbox{Region }II (P)\\ 
\end{array} \right.\
\end{eqnarray}
The partial density of states are 
\be
D_{S(P)}(\epsilon) = 2 \times \int_{\vec{p} \in I (II)} (dp) \,\,\delta(\epsilon-\epsilon_{\vec{p}})
\ee
and the self-consistency equation from Eq(\ref{eqn:SC_DCA}) is
\be
G_0 = (G_S + G_P)/2 \nonumber \\
G_1 = (G_S - G_P)/2
\label{eqn:SC_2-site}
\ee
with
\begin{eqnarray}
G_{S(P)} = \int
\frac{ D_{S(P)}(\epsilon)\,\,d \epsilon}{ \omega + \mu - \epsilon_{\vec{p}} -(\Sigma_0+(-)\Sigma_1)} 
\end{eqnarray}

\subsection{Formalism of the 4-site DCA Method}

In the 4-site DCA the Brillouin zone is divided into four sectors which are labelled as S, P, and D, as shown
in Fig(\ref{fig:k_points})(b). Four $\vec{K}$ points are $(0,0)$ $(\pi,0)$ $(0,\pi)$ $(\pi,\pi)$ leading
to four $\vec{R}$ as $(0,0)$ $(1,0)$ $(0,1)$ $(1,1)$.
The partial DOS is defined as 
\be
D^{(4)}_{S(P,D)}(\epsilon) = 4 \times \int_{\vec{p} \in S (P,D)} (dp) \,\,\delta(\epsilon-\epsilon_{\vec{p}})
\ee
where the superscript (4) is used to distinguish from the partial DOS in 2-site DCA. 
After solving a 4-site impurity cluster problem, in the disordered phase one gets
\be
\hat{G}^{imp} = \left( \begin{array}{cccc}
    G_0 & G_1 & G_2 & G_1   \\  G_1 & G_0 & G_1 & G_2 \\
    G_2 & G_1 & G_0 & G_1   \\  G_1 & G_2 & G_1 & G_0
\end{array}  \right)   \,\,\,
\hat{\Sigma}^{imp} = \left( \begin{array}{cccc}
    \Sigma_0 & \Sigma_1 & \Sigma_2 & \Sigma_1   \\    \Sigma_1 & \Sigma_0 & \Sigma_1 & \Sigma_2 \\
    \Sigma_2 & \Sigma_1 & \Sigma_0 & \Sigma_1   \\    \Sigma_1 & \Sigma_2 & \Sigma_1 & \Sigma_0
\end{array}  \right)
\ee
and the momentum-dependent self energies are
\be
\Sigma_S &=& \Sigma_0 + 2 \Sigma_1 + \Sigma_2 \nonumber \\
\Sigma_P &=& \Sigma_0 - \Sigma_2 \nonumber \\
\Sigma_D &=& \Sigma_0 - 2 \Sigma_1 + \Sigma_2  
\ee
and correspondingly the components of lattice Green's functions are
\be 
G_{S(P,D)} = \int \frac{ D_{S (P,D)}(\epsilon) d\epsilon  }{ i \omega_n + \mu - \epsilon - \Sigma_{S (P,D)} }
\ee
The self-consistency equations are 
\be
G_0 &=&  (G_S+2G_P + G_D)/4\nonumber \\
G_1 &=&  (G_S- G_D)/4\nonumber \\
G_2 &=&  (G_S-2G_P + G_D)/4
\ee


\begin{figure}[htbp]
\begin{center}
   \epsfig{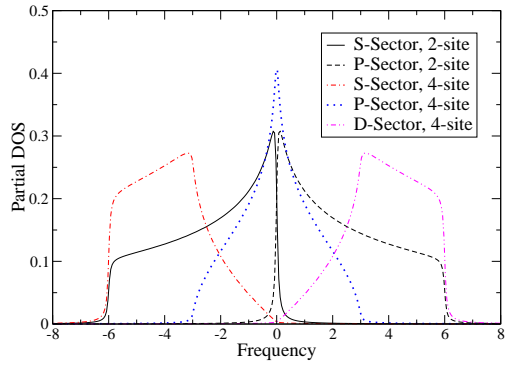}
   \caption{(Color online) The PDOS for 2-site and 4-site DCA partitioning on the square lattice
   	with nearest neighbor hopping. The total bandwidth is 12.
	}
   \label{fig:PDOS_2and4}
\end{center}
\end{figure}

\subsection{The band structure and partial density of states}
\begin{figure}[htbp]
\begin{center}
   \epsfig{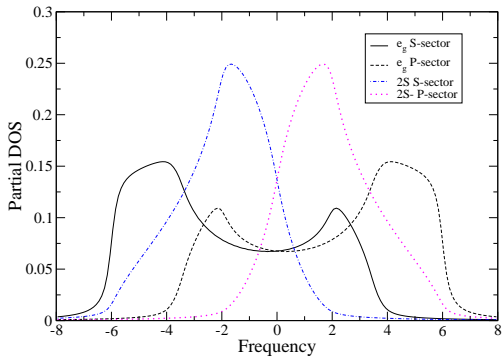}
   \caption{(Color online) Partial DOS for 3D 2$S$ and $e_g$ bands. 
	}
   \label{fig:PDOS_3D}
\end{center}
\end{figure}

Throughout this paper we will consider the square and cubic lattices with only the nearest neighbor hopping
and with total bandwidth chosen to be 12.
For the square lattice with one $S$-like orbital per site, the band dispersion is
\be
\epsilon_{\vec{k}} = -2t (\cos(k_x)+ \cos(k_y))
\ee
The corresponding PDOS for 2-site and 4-site DCA with bandwidth 12 ($t=1.5$ in this case) 
are shown in Fig(\ref{fig:PDOS_2and4}).
For the cubic lattice with one $S$-like orbital per site, the band dispersion is
\be
\epsilon_{\vec{k}} = -2t (\cos(k_x)+ \cos(k_y)+ \cos(k_z))
\label{eqn:Ek_2S}
\ee
The manganite material \cite{CMR} are characterized by a pseudocubic lattice but because
the important orbitals are transition metal $e_g$ orbitals for which the hopping is direction dependent. 
The band dispersion is described by a matrix as 
\be
\hat{\epsilon}_{\vec{k}} = \epsilon_0 \hat{e} + \epsilon_z \hat{\tau}_z +  \epsilon_x\hat{\tau}_x  
\label{eqn:Ek_eg}
\ee
where $\hat{e}$, $\hat{\tau}_x$, and $\hat{\tau}_z$ are unit and Pauli matrices;
$\epsilon_0 = -t (\cos(k_x)+ \cos(k_y)+ \cos(k_z))$, $\epsilon_z = -t[ \cos(k_z) - (\cos(k_x)+ \cos(k_y))/2]$, and
$\epsilon_x = -t \frac{\sqrt{3}}{2}(\cos(k_x)- \cos(k_y))$.
The PDOS for 2-site cubic lattice are plotted in Fig(\ref{fig:PDOS_3D}).

\section{The polaron problem and its impurity cluster}
\subsection{Overview}
In this section we briefly review the breathing mode polaron model and its solution by
dynamical mean field theory. Based on the properties of this model, we provide two
characterizations of the importance of short-ranged correlation.

{\it The breathing mode coupling:}
The Hamiltonian for the polaron model is
\be
H = \sum_{\langle ij \rangle, \alpha \beta} t_{ij} c^{\dagger}_{i \alpha} c_{j \beta}  + 
\sum_i \left[ Q_i n_i + \frac{Q_i^2}{2U}  \right]
\label{eqn:H_B}
\ee
where $\alpha$, $\beta$ can be spin or orbital indices, $Q_i$ is the local breathing mode distortion
with $U$ the restoring energy, and $n_i = \sum_{\alpha } c^{\dagger}_{i \alpha} c_{i \alpha} $ is the local charge density. 
This model has been solved using S-DMFT in Ref\cite{Millis_96}. The main physical effect of the breathing mode $Q_i$
is to trap electrons. In particular when the coupling $U$ exceeds a density dependent critical $U_c(n)$
the system becomes insulating at zero temperature.  The physical interpretation of
the insulating behavior is as follows: 
because a site with a higher density induces a lattice distortion which binds the electron and lowers the energy,
the system eventually  is phase separated into empty and maximally occupied sites where
at empty sites there is no lattice distortion while at maximally occupied site the local distortion
is at its strongest value. The insulating behavior is a consequence of the classical phonon approximation.
Quantum fluctuations of phonons would lead to a small but non-zero conductivity at non-rational $N$.

\subsection{Solution by dynamical mean field theory}

{\it Local partition function:}
In any cluster DMFT calculation, the impurity cluster problem is defined by a partition function. 
For the polaron model on an $N_s$ site cluster, the local distortion fields
are described by $N_s$ classical fields $Q_i$ which couple
to a local density which is quadratic in fermionic operators. In this case, the fermionic 
degrees of freedom can be integrated out and the resulting partition function is 
an integral over $N_s$ classical fields 
\be
Z^{imp,N_s} = \int_{-\infty}^{\infty} \Pi_{i=1}^{N_s} dQ_i e^{-V(\{Q_i\})/T}
\ee
with 
\be
V(\{Q_i\}) =   - 2 T \times \mbox{Tr} \log [\hat{a}-\mbox{diag}(Q_1,Q_2,...,Q_{N_s})]
 +\frac{1}{2 U} \sum_{i=1}^{N_s} Q_i^2
\ee
Here, $\hat{a}$ is a $N_s \times N_s$ matrix composed of a set of Weiss functions $\{a_i\}$.
The 2 in front of the trace accounts for two degenerate local orbitals.

For a 2-site cluster in the disordered phase, two sets of Weiss functions $a_0, a_1$ are needed
and 
\be
\hat{a}^{2-site} =   \bbmatrix{cc} a_0  & a_1 \\ a_1 & a_0   \eematrix 
\ee
For a 4-site cluster, three sets of Weiss functions $a_0, a_1, a_2$ are required and 
\be
\hat{a}^{4-site} = \bbmatrix{cccc} 
a_0  & a_1 & a_2 & a_1 \\ a_1 & a_0  & a_1 & a_2 \\ 
a_2 & a_1 & a_0  & a_1  \\ a_1 & a_2&a_1 & a_0 \eematrix
\ee
The Weiss function, Green's function and self energies in real and imaginary axes are 
calculated using the procedure described in Ref\cite{Millis_96, Okamoto_05}. 
The main quantity we will show is the spectral function $A(\omega)$. 
The zero in frequency $\omega$ is the Fermi energy. We define metal and insulator by the absence or 
presence of a gap at the Fermi level.
To facilitate later comparisons, the total bandwidth is fixed at 12 for all band structure considered
in this paper. 

{\it Defining the importance of short-ranged effects:}
How to determine the degree of short ranged correlation is somehow arbitrary. Here we provide a definition
in the context of cluster dynamical mean field theory: because the single-site DMFT completely
neglects the spatial correlation, we define the strength of the short ranged effect from the $deviation$ 
of cluster results from single-site DMFT results. Two quantitative measurements are now stated. First, 
both single site and cluster C-DMFT require a critical (minimum) interaction strength,
$U_c^{1-site}$ and $U_c^{cluster}$, to get the insulating phase at zero temperature, and the ratio
$U_c^{cluster}/U_c^{1-site}$ ($<1$ generally) is a measure of the importance of short ranged correlations.
Second one can calculate the intersite phonon correlation function $\langle Q_iQ_j \rangle$ 
($\equiv \int_{-\infty}^{\infty} \Pi dQ_k \,\,[Q_iQ_j]\,\, e^{-V( \{Q_k\})/T}/Z^{imp}$).
In single-site DMFT, $\langle Q_iQ_j \rangle = \langle Q_i \rangle \langle Q_j \rangle = 0 $  
while in cluster DMFT $\langle Q_iQ_j \rangle \neq 0 \rangle$. Since that the number
of intersite correlation functions increases exponentially when going to larger cluster size
(including multi-site correlations), we shall primarily use the reduction of $U_c$ in this paper.

\subsection{Pole in self energy}
We shall see that in the models we solve the insulating phases are associated with pole-like structures
in the self energy
\be
\Sigma \sim \frac{V^2}{\omega - \Delta - i \gamma}
\label{eqn:pole}
\ee
characterized by amplitude $V^2$, position $\Delta$, and damping $\gamma$ which is small or zero.
The presence of a pole is related to insulating behavior because insulating
behavior is characterized by a gap, i.e. a region around the Fermi level (which we take to be 0)
around which $Im [G(\vec{k},\omega)] = 0$. If $Im[G]=0$ then $Im[\Sigma] = 0$ except at poles.

A gap in the spectral function at $\omega$ implies both that $Im[\Sigma(\omega)]$ and that there 
is no solution to $Re[G^{-1}(\vec{k},\omega)] = 0$. In the DCA approximation this equation may be written as
\be
\omega - \epsilon_{\vec{k}} - Re[\Sigma_{\vec{K}_i}(\omega)] = 0
\label{eqn:G_inverse}
\ee
for each momentum sector $\vec{K}_i$ and for $\epsilon_{\vec{k}}$ in the allowed range
corresponding to sector $\vec{K}_i$. If the main contribution of the self energy is the pole,
then the boundaries of the gap region can be obtained from the roots of 
$ \omega - \epsilon - V^2/(\omega-\Delta) = 0$ which are
\be
\omega_{+(-)} = \frac{1}{2} [(\epsilon+\Delta) +(-) \sqrt{(\epsilon-\Delta)^2+4V^2}]
\label{eqn:roots}
\ee

Clearly a near Fermi level pole in $\Sigma$
implies a large magnitude of $Re[\Sigma]$ at low $\omega$, meaning that there is no 
solution to Eq(\ref{eqn:G_inverse}) for the allowed range of $\epsilon_{\vec{k}}$.
There are two interesting cases. If we consider density wave order at wavevector $\vec{Q}$
and neglect the long-ranged order by focusing on the diagonal elements of $G$, we find
\be
\Sigma_{DW}(\vec{p},\omega) \sim \frac{V^2}{\omega-\epsilon_{\vec{p}+\vec{Q}}}
\label{eqn:Sigma_DW}
\ee
(A broadened version of $\Sigma_{DW}$ is the basis of the Lee-Rice-Anderson theory
of the pseudogap in thermally disordered charge density waves).

$\Sigma_{DW}$ expresses the physics of level repulsion: if $\epsilon_{\vec{p}+\vec{Q}}>\epsilon_{\vec{p}}$
then the presence of $\Sigma_{DW}$ pushes the solution of Eq(\ref{eqn:G_inverse})
(which would be $\omega=\epsilon_{\vec{p}}$) to an energy lower than $\epsilon_{\vec{p}}$, while
also creates a second solution at $\omega>\epsilon_{\vec{p}+\vec{Q}}$. We shall see that
in many cases the cluster DMFT theories, which replace self energies by some sort of
momentum averaged self energies, lead to just this physics, with the self energy corresponding to
a $\vec{K}$ sector with $\epsilon_{\vec{k}}$ dominantly less than 0 having a pole at an energy $\Delta>0$,
so that the effect of $\Sigma$ is to push the states in this $\vec{K}$ sector away from $\omega \sim 0$.
This self energy arises from near resonant coupling to states in the complementary $\vec{K}$ sector.
We interpret this as the density wave (Slater-like) mechanism. 

On the other hand, one sometimes finds cases where the self energy corresponding to a given
$\vec{K}$ sector has a pole at an energy $\Delta$ in the middle of the allowed energy range arising
from coupling to all $\vec{K}$ sectors. In that sector, the pole in self energy pushes states away
in both directions. We identify this as the Mott mechanism. 

{\it Extracting pole parameters:} The pole parameters -- the position, the amplitude, and the damping -- from 
$\Sigma \sim V^2/(\omega-\Delta + i\gamma)$ are obtained as follows. 
The position $\Delta$ is estimated by $Re[\Sigma] = 0$ (any Hartree contribution is absorbed into
the chemical potential, which we define to be 0); the amplitude $V^2$ and
damping $\delta$ are determined by fitting the $Re[\Sigma]$ to
$V^2(\omega-\Delta)/[(\omega-\Delta)^2+\gamma^2]$.

\section{The half-filled case}
\subsection{Overview}
In this section the DCA method with cluster size 1, 2, 4 is used to analyze the two dimensional square 
lattice with nearest neighbor hopping $t=1.5$. The model displays a metal insulator transition
as the interaction $U$ is increased beyond a critical value $U_c$, which as shown in Table I is in the
range of $3\sim 5$ depending on the cluster size. 
\begin{center}
\begin{tabular}{l|l|l|l} 
cluster size    & 1-site   &  2-site & 4-site \\ \hline
$U_c$  & $\sim 5$ &  $\sim 3$ & $3\sim 3.5$  
\end{tabular} \\ 
Table I: Gap opening $U_c$ for the polaron model on a square lattice with bandwidth 12
 obtained from DCA with different cluster size
\end{center}

The excitation spectra are obtained for different cluster sizes.
The physics of the metal-insulator transition and the variations in results with
cluster size are related to the behavior of poles in the self energy. In the rest of this section 
we present and analyze the results 
for 1-site, 2-site and 4-site DCA calculation for the local interaction range $U \lesssim U_c^{1-site}$ 
($U_c^{1-site}$ is the gap opening $U$ for single-site DMFT calculation).

\subsection{Single-site results}
\begin{figure}[htbp]
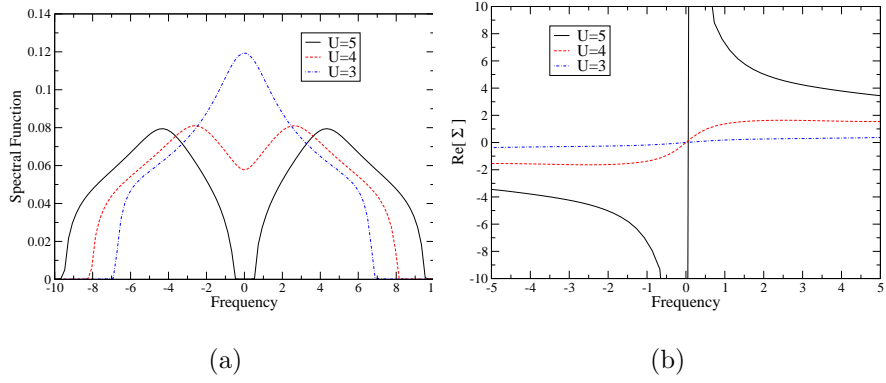

\begin{center}
   \subfigure[]{\epsfig{file = DOS_t1.5_t1_0.00_U3-5_Occu1.0_Temp0.1.eps,  width = 0.35\textwidth}}
   \subfigure[]{\epsfig{file = Sigma_SDMFT_t1.5_t1_0.00_U3-5_Occu1.0_Temp0.1.eps,  width = 0.35\textwidth}}
   \caption{(Color online) Results from single-site DMFT calculation for two-dimensional square lattice with nearest neighbor 
   	hopping $t=1.5$ (bandwidth 12), polaronic coupling (Eq(\ref{eqn:H_B})) $U=3,4,5$ and $T=0.1$:
   	(a) spectral functions; (b) real part of self energies. 
	}
   \label{fig:1siteDCA_Sigma_G}
\end{center}
\end{figure}
We first present the single-site DMFT calculation. Fig(\ref{fig:1siteDCA_Sigma_G})(a) shows the spectral functions for
$U=3,4,5$ at $T=0.1$. The gap-opening interaction for the single-site calculation, $U_c^{1-site}$, is 
slightly less than 5. As discussed in the previous section, the insulating behavior is related to a pole
in the self energy. This is shown in Fig(\ref{fig:1siteDCA_Sigma_G})(b) which reveals at $U=3$ the real part of the 
self energy is very weak; at $U=4$ a strongly broadened pole begins to appear, and at $U=5$ an undamped pole occurs.

The self energy sum rule for this model is $\int\frac{d \omega}{\pi} Im[\Sigma(\omega)] = U^2$.
By fitting $\Sigma$ (not shown here) at $U=5$ we find the pole strength $V^2 = 7.5$ so that 
 the pole contains roughly $30\%$ of the total self energy sum rule. As 
$U$ is increased the fraction of the self energy sum rule contained in the pole increases to $90\%$
of the total at $U=10$ and $99\%$ at $U=20$. However even for $U=5$ where there is
a substantial non-pole contribution to $Im[\Sigma]$, the size of the gap is to a good approximation
given by inserting $\Sigma \sim 7.5/\omega $ into the quasi-particle equation Eq(\ref{eqn:G_inverse}).

\subsection{2-site results}

\begin{figure}[htbp]
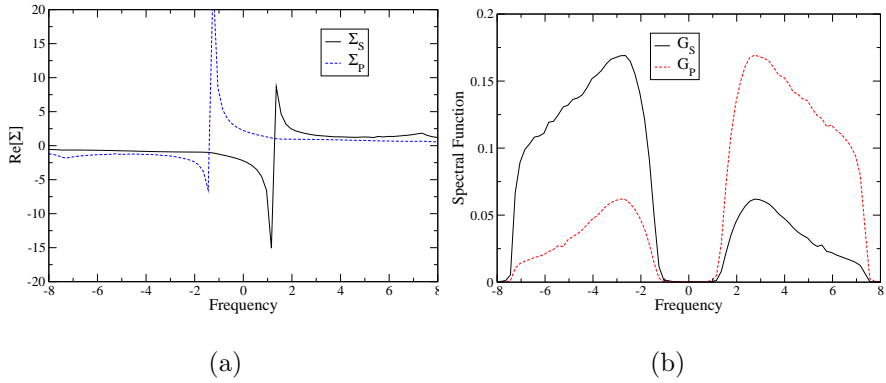

\begin{center}
   \subfigure[]{\epsfig{file = Sigma_2SiteDCA_t1.5_U4.00_Occu1.0_Temp0.100.eps,  width = 0.35\textwidth}}
   \subfigure[]{\epsfig{file = DOS_2site_t1.5_t1_0.000_U4.00_Occu1.0_Temp0.100_update0.96_accu_5.0E-3.eps,  width = 0.35\textwidth}}
   \caption{(Color online) Results from 2-site DCA calculation for two-dimensional square lattice with nearest neighbor 
   	hopping $t=1.5$ (bandwidth 12), polaronic coupling (Eq(\ref{eqn:H_B})) $U=4$ and $T=0.1$.
   	(a) Real part of self energies for $S$ and $P$ sectors. (b) Partial spectral functions  for $S$ and $P$ sectors.
	}
   \label{fig:2siteDCA_Sigma_G}
\end{center}
\end{figure}
In this subsection we show results from 2-site DCA calculation. We focus on 
$U=4$ where the solution is insulating but in the 1-site approximation would be metallic. 
Fig(\ref{fig:2siteDCA_Sigma_G})(a) shows the real part of the self energy for $S$ and $P$ sectors at
$U=4$ $T=0.1$ in for 2-site DCA method. The most prominent feature is poles. 
Comparison to Fig(\ref{fig:PDOS_2and4}) shows that the pole in the $S$ sector (here $\Delta \sim 1.5$)
lies above the energy range where the $S$ sector PDOS is nonvanishing (Fig(\ref{fig:PDOS_2and4}));
while pole in the $P$ sector lies below the support of the $P$ sector PDOS. 
As discussed previously, the pole pushes the main part of the $S$ sector PDOS down in
energy while generating some excitation at energies above the pole energy. The opposite effect
occurs in the $P$ sector. Therefore in the 2-site DCA the origin of the gap opening transition
is a coarse-grained (in momentum space) version of the Slater physics.

\subsection{4-site results}
\begin{figure}[htbp]
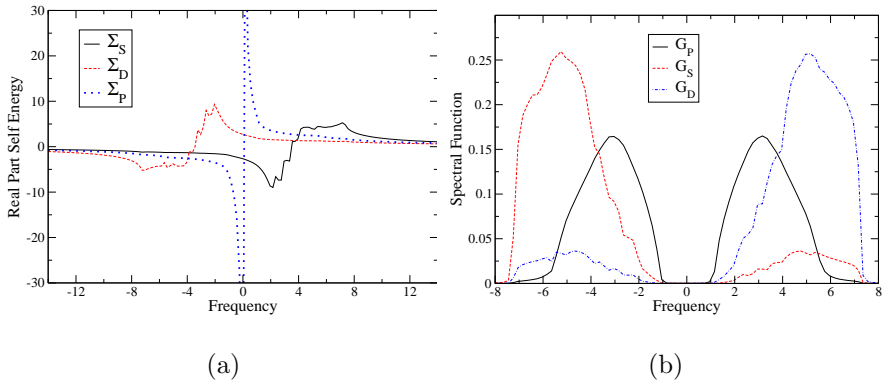

\begin{center}
   \subfigure[]{\epsfig{file = SigmaReal_4SDCA_SDMFT_t1.5_U4.00_Occu1.0_Temp0.100.eps,  width = 0.35\textwidth}}
   \subfigure[]{\epsfig{file = DOS_4site_t1.5_t1_0.000_U4.00_Occu1.0_Temp0.100_update0.98_accu_10.0E-3.eps, width = 0.35\textwidth}}
   \caption{(Color online) Results from 4-site DCA calculation for two-dimensional square lattice with nearest neighbor 
   	hopping $t=1.5$ (bandwidth 12), polaronic coupling (Eq(\ref{eqn:H_B})) $U=4$ and $T=0.1$.
   	(a) Real part of self energies for $S$ $P$ and $D$ sectors. (b) Partial spectral functions for $S$ $P$ and $D$ sectors.
	}
   \label{fig:4siteDCA_Sigma_G}
\end{center}
\end{figure}

As in the previous subsection we focus on the insulating solution. 
In the 4-site DCA calculation, there are three different sectors $S$ $P$ and $D$ in momentum space.
Their self energies and partial spectral functions are shown in Fig(\ref{fig:4siteDCA_Sigma_G}). As seen from  Fig(\ref{fig:PDOS_2and4}), the PDOS of $S$ and $D$ sectors barely touch each other and the poles obtained from 4-site
DCA have exactly the same effect as the $S$ and $P$ sectors in the 2-site calculation --
$S-$ and $D-$ sectors repel each other. However in the $P$ sector
the pole  energy is zero and the physics is "Mott-like". 
We emphasize that in DCA equation all three sectors are coupled so states in the 
$P$ sector are not isolated from states of other two sectors. 
Thus while the $P$ sector bandwidth is a factor of two smaller than the total bandwidth, the 4-site
$U_c^{4-DCA} \sim 3.5$ is about $70\%$ of the single site value. 

We observe that the polaron model (local coupling $Q \hat{n} + \frac{1}{2} Q^2/U$, Eq(\ref{eqn:H_B})) 
is in this respect different form the Hubbard model, where $U_c^{4-site} \sim 0.4 U_c^{1-site}$
\cite{Gull_08, Park_08}. We further observe that the $U$-driven transition in the 4-site DCA
approximation to the polaron model is continuous while in the Hubbard model the transition
is strongly first order. In the Hubbard model the transition was associated with the dominance
of a particular plaquette singlet state \cite{Gull_08} not important here, suggesting that the smaller
$U_c^{4-site}$ in the Hubbard model case is a manifestation of the strong short-ranged ordering effect.

\subsection{Comparison between 2-and 4-site results}
\begin{figure}[htbp]
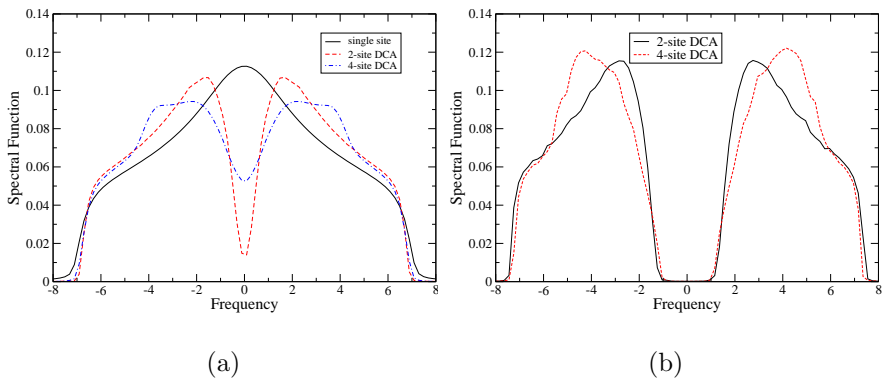

\begin{center}
\subfigure[]{\epsfig{file = DOS_t1.5_U3.00_Occu1.0_Temp0.2_all.eps, width = 0.35\textwidth}}
\subfigure[]{\epsfig{file = DOS_t1.5_U4.00_Occu1.0_Temp0.1_all.eps, width = 0.35\textwidth}}
\caption{(Color online) Spectral functions calculated by 2-site and 4-site DCA approximations for the
	two-dimensional square lattice with nearest neighbor hopping $t=1.5$, polaronic coupling (Eq(\ref{eqn:H_B}))
	$U=3$, $T=0.2$ (a) and $U=4$, $T=0.1$(b). 
	}
   \label{fig:DOS_DCA2_and4}
\end{center}
\end{figure}

\begin{figure}[htbp]
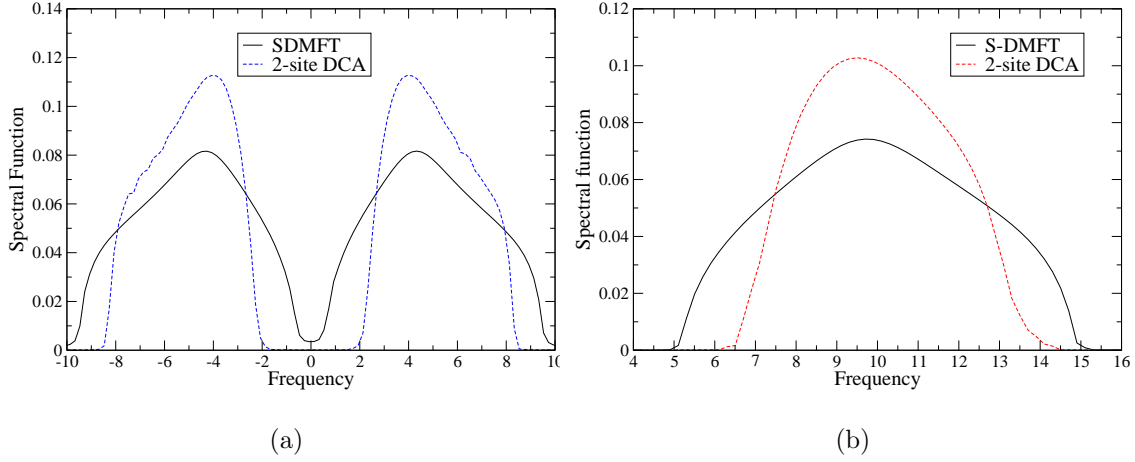

\begin{center}
   \subfigure[]{\epsfig{file = DOS_2site_SDMFT_t1.5_t1_0.000_U5.00_Occu1.0_Temp0.100.eps,  width = 0.45\textwidth}}
   \subfigure[]{\epsfig{file = DOS_Upper_1_2site_t1.5_t1_0.000_U10.00_Occu1.0_Temp0.1.eps,  width = 0.45\textwidth}}
   \caption{(Color online) Spectral functions from S-DMFT and 2-site DCA calculation for the
	two-dimensional square lattice with nearest neighbor hopping $t=1.5$, polaronic coupling $U=5$ (a) 
	and $U=10$ (b) , $T=0.1$ and $N=1$. For $U=10$ only the upper Hubbard band is shown.
	}
   \label{fig:2siteDCA_Sigma_G_U5N1}
\end{center}
\end{figure}
Now we compare results for 2-site and 4-site DCA calculations in Fig(\ref{fig:DOS_DCA2_and4}).
At $U=2$ (not shown) the spectral functions are similar. At $U=3$ ($\sim U_c^{2-site}$ but $<U_c^{4-site}$)
the spectra are different; but by $U=4$ the spectra have again become very similar.
As $U$ continues to increase the difference between 2 and 4-site calculations remain
small, but both remain different from the results of the single site calculation:
in the multi-site case the gaps are larger and the gap edge more sharply defined.
This is a consequence of short-ranged order (see also Table II, third row) and 
is demonstrated in Fig(\ref{fig:2siteDCA_Sigma_G_U5N1}) in which the results 
obtained from single-site DMFT and 2-site DCA are shown for $U=5$ and $U=10$.
One sees that the sharp band edge remains at large $U$.

\section{$T$ and $U$ dependence}

In this section we discuss the temperature and interaction dependence of the poles in the self energies.
As previously discussed, in both single-site and cluster DMFT calculation, 
the insulating behavior can be traced back to a pole-like
structure in self energy. The pole is characterized by its location $\Delta$, amplitude $V^2$, and damping $\delta$ 
which we now analyze for different interaction strengths $U$ and temperature $T$.

\begin{figure}[htbp]
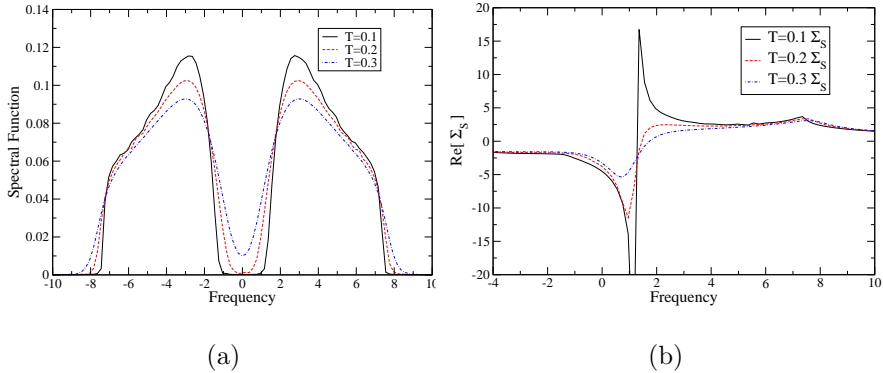

\begin{center}
   \subfigure[]{\epsfig{file = DOS_2site_t1.5_t1_0.000_U4.00_Occu1.0_Temp0.1-0.5.eps,  width = 0.35\textwidth}}
   \subfigure[]{\epsfig{file = Sigma_s_2SiteDCA_t1.5_U4.00_Occu1.0_Temp0.1-0.3.eps, width = 0.35\textwidth}}
   \caption{(Color online) Results from 2-site DCA calculation for the two-dimensional square lattice with nearest neighbor 
   	hopping $t=1.5$ (bandwidth 12), polaronic coupling $U=4$ and $T=0.1$-0.3.
   	(a) Spectral functions. (b) Real part of self energies for $S$ sectors. 
	}
   \label{fig:2site_T}
\end{center}
\end{figure}

Fig(\ref{fig:2site_T}) shows the temperature dependence of the density of states
and the real part of the $S$ sector self energies calculated in the 2-site DCA approximation
for $U=4$ (bandwidth 12). As the temperature is raised, the gap begins to fill in, and this
change is accompanied by an obvious decrease in the size and sharpness of the pole structure. 
While the pole is not well separated from the other contributions to $\Sigma$, a simple fit
yields a gradual decrease in the amplitude $V^2$ (Eq(\ref{eqn:pole})) from 13 at $T=0.1$ to 9 at $T=0.3$, and
a rapid increase in the damping $\gamma$ from negligible at $T=0.1$ to $\gamma=0.3$ at $T=0.2$ 
to $\gamma=0.6$ at $T=0.3$. The shift in pole position is very small. Thus the main contribution
to the destruction of the insulating phase is a broadening of the pole.
A similar conclusion is found in the 4-site calculation (not shown here). The $T$-independence
of the pole position is in contrast to the results of Ref\cite{Fuhrmann_07} who found the pole position
was temperature dependent, tracking the cluster spin correlation function. We believe this is
a difference between the strong coupling limit considered in Ref\cite{Fuhrmann_07} and the
intermediate coupling studied here.

We next discuss the $U$ dependence in DCA calculation.  We begin with the strong coupling limit
in which the excitations are expected to be roughly $ +(-) U$. 
From Eq(\ref{eqn:roots}), when amplitude of the pole is very large compared to both the pole location
$\Delta$ and the half-bandwidth of non-interacting spectrum $\Lambda$, the excitations are roughly
$\omega_{+(-)} \sim +(-) V$. Therefore in this limit one expects the pole amplitude is $U^2$ and 
the location of the pole is well within the gap, i.e. $|\Delta|<<U$ (see the first row of Table II).
As $U$ is decreased, the bandgap and the pole energies both decrease, but the bandgap decreases much
faster so that for the intermediate $U$ the pole in the self energy lie actually within the
continuum of excited states (see the first and second row of Table II), so is technically not a pole but is a resonance.
Thus a pole in the self energy is not necessary for insulating behavior. We also find that the
"pole" energy does not have a simple relation to the intersite phonon correlation
$\langle Q_1 Q_2 \rangle$ (third row of Table II). This is again in contrast to results of Ref\cite{Fuhrmann_07}, which
found via an approximate analytical calculation that in the Hubbard model the
pole energy was proportional to the spin correlation function $\langle S_1 \cdot S_2\rangle$.
We believe the difference arises from the coupling strength.
The $U$ dependence is summarized in Table II whose first row gives the $S$ sector pole position
obtained from 2-site DCA calculation for different $U$ at $T=0.1$, second row the gap edge, third row
the intersite correlation function.
\begin{center}
\begin{tabular}{|l|l|l|l||l|l|} \hline
$U$ & 3 & 4 & 5  & 10 &15\\ \hline
$S-$ pole location   & 1.21   & 1.31  & 1.51  & 2 & 2 \\ \hline
gap edge & $\sim 0$ & 1 & 1.74 & 6.75 & 11 \\ \hline
$-\langle Q_1 Q_2 \rangle/U^2$  & 0.40   & 0.61  & 0.73  & 0.92 & 0.924  \\ \hline
\end{tabular} \\
Table II: the pole position, gap edge, and intersite phonon correlation for different polaronic
coupling $U$ obtained from 2-site DCA calculation on the two dimensional square lattice
with bandwidth 12. \\
\end{center}



\section{Filling and cluster size}
\subsection{Overview}
In this section we discuss band filling effects in cluster DMFT theory. The basic finding is that the filling
determines the minimum cluster size required to provide a significant difference
between single-site and cluster DMFT results -- the minimum size
increases as the system is doped from the half filling. In this section we restrict ourselves
to the square lattice only. 

\subsection{General filling 2-site }


\begin{figure}[htbp]
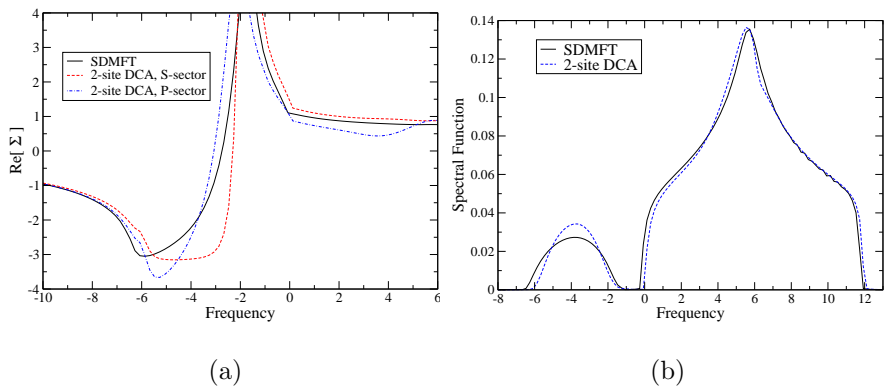

\begin{center}
   \subfigure[]{\epsfig{file = Sigma_2SiteDCA_SDMFT_t1.5_t1_0.000_U5.00_Occu0.2_Temp0.1.eps,  width = 0.35\textwidth}}
   \subfigure[]{\epsfig{file = DOS_2site_SDMFT_t1.5_t1_0.000_U5.00_Occu0.2_Temp0.1.eps,  width = 0.35\textwidth}}
   \caption{(Color online) Results from single-site and 2-site DCA calculations 
   for the square lattice with nearest neighbor hopping
   	$t=1.5$, polaronic coupling $U=5$, $T=0.1$ at $N=0.2$.
   	(a) Real part of self energies for $S$ and $P$ sectors. (b) Spectral functions.
	}
   \label{fig:2siteDCA_Sigma_G_U5N0.2}
\end{center}
\end{figure}

In this subsection we compare the doping dependence found in single-site and 2-site DCA for $U=5$ above
the critical value for insulating behavior in both approximations.
The discussion of Fig(\ref{fig:2siteDCA_Sigma_G_U5N1}) in Sec IV showed the pronounced differences between
single-site and cluster DMFT calculations at density $N=1$. For the very low doping $N=0.2$ on the other hand, 
the single-site DMFT and 2-site DCA behave 
very similarly. Fig(\ref{fig:2siteDCA_Sigma_G_U5N0.2})(a) shows that within DCA poles in 
self energies in the two sectors are almost identical, and are quite close to that in  single-site DMFT. 
Consequently the spectral
functions obtained from both methods are very similar ( Fig(\ref{fig:2siteDCA_Sigma_G_U5N0.2})(b)).
This can be understood as follows. At $N=0.2$, the Fermi energy only lies deep inside the  $S$-sector PDOS,
the $P$ sector is irrelevant, and 
the DCA calculation within the $S$-sector is just like a single-site DMFT calculation.

\subsection{Near quarter filling, 2-site and 4-site }

\begin{figure}[htbp]
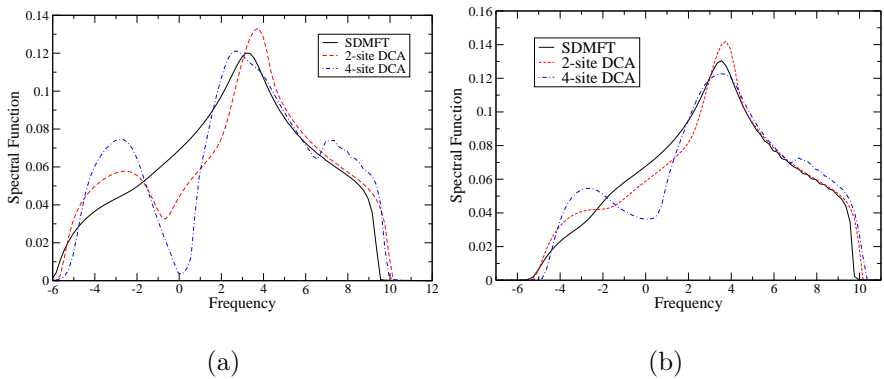

\begin{center}
\subfigure[]{\epsfig{file = DOS_0-2-4site_t1.5_t1_0_U4.00_Occu0.5_Temp0.1.eps, width = 0.35\textwidth}}
\subfigure[]{\epsfig{file = DOS_0-2-4site_t1.5_t1_0_U4.00_Occu0.4_Temp0.1.eps, width = 0.35\textwidth}}
\caption{(Color online) Spectral functions calculated by single-site DMFT, 2-site and 4-site DCA approximations at
	$U=4$, $T=0.1$, $N=0.5$(a) and $N=0.4$(b) . 
	}
   \label{fig:DOS_DCA2_and4_quarter}
\end{center}
\end{figure}

\begin{figure}[htbp]
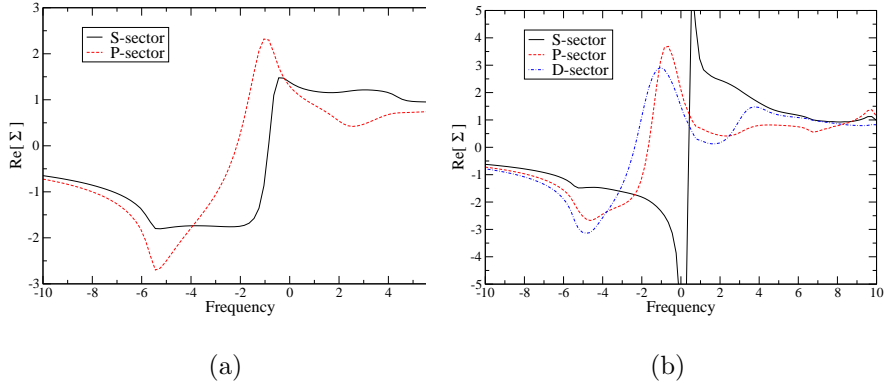

\begin{center}
\subfigure[]{\epsfig{file = Sigma_2SiteDCA_t1.5_t1_0.000_U4.00_Occu0.5_Temp0.1.eps, width = 0.35\textwidth}}
\subfigure[]{\epsfig{file = SigmaReal_4SDCA_t1.5_U4.00_Occu0.5_Temp0.1.eps, width = 0.35\textwidth}}
\caption{(Color online) Self energies calculated by 2-site (a) and 4-site (b) DCA approximations 
	for the two dimensional square lattice with bandwidth 12 at
	$U=4$, $T=0.1$, $N=0.5$. 
	}
   \label{fig:Sigma_DCA2_and4_quarter}
\end{center}
\end{figure}

In the classical polaronic model, insulating behavior can occur at any density if the interaction is
strong enough. In this subsection we show that the Slater mechanism of shifting of states in $k$
space can lead to insulating behavior at smaller $U$, if the band filling is such that
one of the sectors is mostly filled, while the adjacent one with higher energies is nearly empty.
However these cluster size effects diminish rapidly in importance as the density is moved away
from commensurate values. The left panel of Fig(\ref{fig:DOS_DCA2_and4_quarter}) compares the results 
from 1-site, 2-site, 4-site DCA approximations to the two dimensional square lattice with 
nearest neighbor hopping at density $N=0.5$ and intermediate $U$. The four site approximation
produces a clearly insulating behavior, the two site gives a moderate "pseudogap" and the single site
approximation yields results very similar to the non-interacting model. The right panel
of Fig(\ref{fig:DOS_DCA2_and4_quarter}) shows that the differences decreases rapidly as the doping 
is moved from the commensurate value.

To unravel the origin of the insulating behavior we consider the self energies for the 4-site DCA
shown in Fig(\ref{fig:Sigma_DCA2_and4_quarter}). The right panel shows a clear pole in the $S$ sector,
lying at an energy slightly above the chemical potential $\mu$ and substantially above the average 
energy of $S$ sector $\sim \mu-0.93$. This pole gives rise to the Slater
effect of pushing most of $S$ sector states down in energy. Consequently the $P$ and $D$ sector self
energies lie below the $P$ and $D$ PDOS and push these states up, 
leaving the gap observed in Fig(\ref{fig:DOS_DCA2_and4_quarter}).

By contrast the left panel of Fig(\ref{fig:Sigma_DCA2_and4_quarter}) shows for the 2-site DCA an 
interesting hybrid behavior. The $P$ sector pole lies below the $P$-sector PDOS and pushes these states up.
The $S$ sector pole lies roughly at $\mu-1$, below the average energy of $S$ PDOS 
$\langle \epsilon \rangle_S^{2-site} = \mu+0.95$  and acts in a Mott-like fashion, 
splitting the band into two.

\section{The role of partial density of state}
In this section we establish the relation between the PDOS defined in Eq(\ref{eqn:Def_PDOS}) and 
the differences between cluster and single-site DMFT calculations. In essence, the more overlap between PDOS,
the more similar are the single-site and cluster calculations.  
To measure the overlap between PDOS of two sectors, we first compute the overlap
$O_{ij} \equiv \int \,d\epsilon D_i(\epsilon) D_j(\epsilon)$. 
One can also use the first two moments of the partial DOS. For PDOS $i$, 
$\langle \epsilon \rangle_i = \int \, d \epsilon \epsilon D_i(\epsilon)$ gives the average
position of the partial DOS while $\sigma_i = \sqrt{\langle \epsilon^2 \rangle_i-\langle \epsilon \rangle^2_i}$ 
tells how spread the PDOS is. The overlap is related to $ \langle \epsilon \rangle_i / \sigma_i$.
We consider two examples. First we compare 2-site DCA and 2-site cellular DMFT solutions to
the polaron model on the square lattice. Second, we compare the approximations to two models involving 
the polaron model defined on a cubic lattice: one has two degenerate $S$-like orbitals per site
the other two $e_g$ orbitals, with dispersions given in Eq(\ref{eqn:Ek_2S}) and Eq(\ref{eqn:Ek_eg}) respectively.

\subsection{Difference between 2-site DCA and 2-site CDMFT for 2D square lattice}

\begin{figure}[htbp]
\begin{center}
   \epsfig{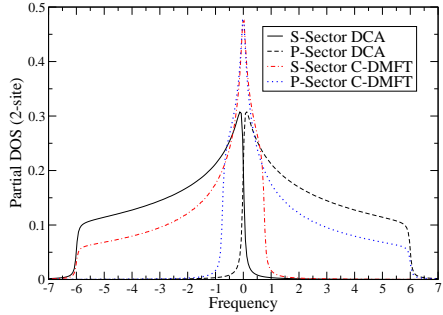}
   \caption{(Color online) Partial DOS for 2-site DCA and Cellular-DMFT on the square lattice with nearest neighbor hopping
   	(
   	$t=1.5$ (bandwidth 12). $U_c$ for 1-site DMFT, 2-site cellular DMFT, 
	2-site DCA approximations are roughly 5, 4, 3 respectively.
	}
   \label{fig:DCAvsCDMFT}
\end{center}
\end{figure}
In this subsection we compare the two site DCA and cellular DMFT solutions to the polaron model
defined on the square lattice. As discussed in Sec II and Appendix A the mean field equations may be
expressed in the same form as
\be
G^{imp}_{S,P} = \int d\epsilon \frac{D_{S,P}(\epsilon)}{\omega - \epsilon - \Sigma_{S,P}(\omega)}
\label{eqn:SC_simplified}
\ee
The two approximations are distinguished by the differences in PDOS, shown in Fig(\ref{fig:DCAvsCDMFT}).
In the DCA case two PDOS do not overlap (the small apparent overlap is the result of numerical broadening);
in the cellular DMFT the overlap is seen to be more substantial. Quantitative information
on the overlaps is given in Table III
\begin{center}
\begin{tabular}{l|l|l|l} 
 & $O_{SP}$ & $\langle \epsilon \rangle_S$ = - $\langle \epsilon \rangle_P$&  $\sigma_S$=$\sigma_P$ \\ \hline
DCA   & 0  & -2.41   & 1.82           \\ \hline
Cellular-DMFT & 0.15  & -1.48   &  1.89    
\end{tabular} \\ Table III: The overlap information of PDOS of $S$ and $P$ sectors for DCA and cellular DMFT
on the square lattice with bandwidth 12. 
$\langle \epsilon \rangle_{S(P)}$ and  $\sigma_S$=$\sigma_P$ are the  average energy 
and the standard deviation of $S(P)$ PDOS. 
\end{center}

\begin{figure}[htbp]
\begin{center}
   \epsfig{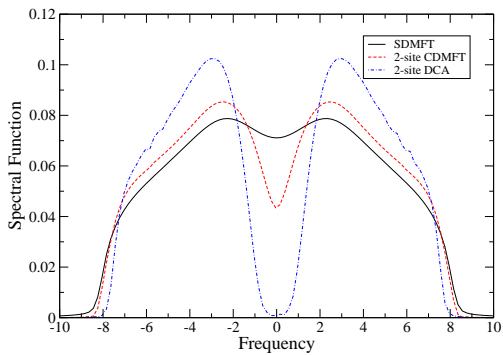}
   \caption{(Color online) 2-site cellular-DMFT and DCA spectral functions calculated for the square lattice
   	with nearest neighbor hopping $t=1.5$ at half-filling, polaronic coupling $U=4$, $T=0.2$. 
   The DCA result is more insulating (having smaller density of state around the Fermi energy) 
   than cellular-DMFT.
	}
   \label{fig:DCAvsCDMFT_sameU}
\end{center}
\end{figure}

To further explore the difference we show in Fig(\ref{fig:DCAvsCDMFT_sameU}) the spectral functions
calculated at $U=4$ and at the relatively high $T=0.2$. The DCA calculation reveals a well defined gap,
whereas the cellular DMFT reveals a weak pseudogap which will evolve to a small true gap as 
$T \rightarrow 0$. As seen from Eq(\ref{eqn:SC_simplified}) these differences can arise 
mathematically only from the difference in PDOS.

Solving the equations reveals  that the gap-opening $U$ of single-site
DMFT at half filling is $U^{1-site}_c \sim 5$ (total bandwidth is 12), for 2-site cellular DMFT 
$U_c^{2-Cellular} \sim 4$ and for 2-site DCA, $U_c^{2-DCA} \sim 3$. 
The difference shows that the DCA expresses short ranged correlations more
strongly than the cellular DMFT.

\subsection{3D $e_g$ and 2$S$ band}

\begin{figure}[htbp]
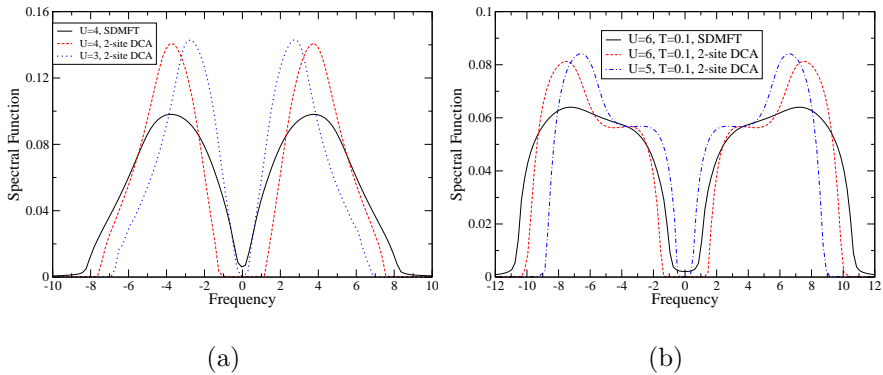

\begin{center}
\subfigure[]{\epsfig{file = DOS_2site_3D_2S_t1.0_U4and3_Occu1.0_Temp0.1.eps,  width = 0.35\textwidth}}
\subfigure[]{\epsfig{file = DOS_DCA3Pi_3D_eg_t2.0_U6and5_Occu1.0_Temp0.1.eps,  width = 0.35\textwidth}}
   \caption{(Color online) Spectral functions for the coupling close to the gap opening $U_c$.
   (a) For two degenerate $S$-band with bandwidth 12, the $U_c^{1-site} \sim 4$ (solid) and $U_c^{2-DCA} \sim 3$ (dashed).
   (b) For $e_g$ band with bandwidth 12, the $U_c^{1-site} \sim 6$ (solid) and $U_c^{2-DCA} \sim 5$ (dashed).
   Those curves are calculated at $T=0.1$. 
	}
   \label{fig:3D_egvs2S}
\end{center}
\end{figure}

In this subsection we compare the results of applying the DCA approximation to two orbitally degenerate 
models defined on a cubic lattice: the $S$-orbital model, with two degenerate bands of dispersion
given in Eq(\ref{eqn:Ek_2S}) and the $e_g$ model, appropriate to the colossal magnetoresistance
manganites, with dispersion given by the eigenvalues of Eq(\ref{eqn:Ek_eg}). 
We choose the bandwidth to be 12 for both bands. The corresponding PDOS are shown in 
Fig(\ref{fig:PDOS_3D}) and the overlaps are given in Table IV.
\begin{center}
\begin{tabular}{l|l|l|l} 
 & $O_{SP}$ & $\langle \epsilon \rangle_S$ (=-$\langle \epsilon \rangle_P$)  &  $\sigma_S$ ($=\sigma_P$) \\ \hline
$2S-$like band & 0.044   & -1.84   & 1.8           \\ \hline
$e_g$ band & 0.052 &-1.82   &  2.98    
\end{tabular} \\  Table IV: The overlap information of PDOS of $S$ and $P$ sectors for $S$ 
(Eq(\ref{eqn:Ek_2S})) and $e_g$ (Eq(\ref{eqn:Ek_eg})) bands on the cubic lattice with bandwidth 12.  
$\langle \epsilon \rangle_{S(P)}$ and  $\sigma_S$=$\sigma_P$ are the  average energy 
and the standard deviation of $S(P)$ PDOS. 
\end{center}
We see that the $e_g$ band has larger PDOS overlap than the $S$ band and 
 thus one expect a smaller short ranged correlation for the $e_g$ case.
Indeed, for 2$S$ band ($U_c^{1-site} \sim 4$) $U_c^{2-DCA}/U_c^{1-site} \sim 3/4 =0.75$ 
while for $e_g$ ($U_c^{1-site} \lesssim 6$) $U_c^{2-DCA}/U_c{1-site} \sim 5/6 =0.83$.
The corresponding spectral functions are provided in Fig(\ref{fig:3D_egvs2S}). We observed that
in the insulating region the DCA leads to a larger gap and sharper band edge than the single-site DMFT,
but the differences are less in the $e_g$ case.

\section{Summary and Discussion}

We first summarize our findings. The cluster-DMFT calculation reduces the critical interaction strength
to open a gap. The magnitude of this is introduced as a quantitative measure of the
importance of short-ranged correlation. 
The partial density of states (PDOS) and the filling of the system are two crucial quantities
in the cluster DMFT calculations. The reduction of critical $U$ in cluster calculations is mainly 
caused by the momentum dependent
pole structure in self energies which pushes PDOS around Fermi energy away from each other.
Similarly we find that significant differences between single-site and cluster calculations only occur when
the Fermi energy involves several PDOS (2 at least). For example in the square lattice, the 2-site
produces small $U$ insulating behavior for half-filling problem but differs little from single-site DMFT 
for quarter-filling. Obtaining a small $U$ insulating state in the quarter filling requires a 4-site cluster.
We show that when the cluster size is large enough, a large separation between PDOS around
Fermi energy implies a strong short-ranged effect.
One notes that the difference between single-site and cluster DMFT results becomes
pronounced for the range of $U$ where $\lesssim U_c^{1-site}$, above which the cluster
calculations lead to a larger gap and sharper bandedge. These features persist
in the strong $U$ limit and are consequences of the short-ranged correlation 
included in the cluster approximation.

The use of cluster DMFT methods to study metal insulator transitions at carrier concentrations different
from half filling is an intriguing issue. Our results suggest that the cluster size must be
tuned according to the filling to be studied. Roughly filling $1/N_s$ requires an $N_s$-site
cluster. Thus for example study of the interesting
phenomena associated with "half-doped" manganites (one carrier for every two two-fold degenerate
Mn orbitals, 1 electron per 4 orbitals, quarter filling), at least a 4-site cluster is required. 
The physics of the metal insulator transitions at $N \neq 1$ is found generally to be Slater-like, 
arising from the separation in energy of $k$ space sectors. 

A generic results is that $N_s$-site ($N_s>1$) cluster DMFT calculations predict smaller critical $U$ for the gap 
opening (metal-insulator) transition than does the single site approximation. Although 
various effects occur, we find that generically the most important role is played by a 
Slater-like effect. The calculation is most naturally viewed in terms of momentum space sectors, and
the dominant contribution to the insulator behavior is given by a pole structures in the
self energy which act to open gaps between the different momentum sectors. 


\section*{Acknowledgment}
We thank Jie Lin, Xin Wang, Hung The Dang for helpful discussion and 
this work is supported by DOE ER 46169.
\appendix

\section{PDOS of cellular DMFT approximation}
In this appendix we derive the partial density of states for 2-site cellular DMFT approximation
for square lattice with nearest neighbor hopping $t$. 
We begin by casting these two methods in the same formalism. 
Once the 2-site impurity cluster problem is solved, one gets Eq(\ref{eqn:G_2}). 
The self consistency equation for both 2-site DCA and cellular-DMFT can be expressed
in the same form similar to Eq(\ref{eqn:CDMFT_SC}) as
\be
\bbmatrix{cc}  G_0 & G_1   \\   G_1 & G_0 \eematrix =
2\times \int_{\vec{k} \in I} (dk)\left[
\omega + \mu - \bbmatrix{cc} \Sigma_0 & \psi_{\vec{k}} +\Sigma_1 \\ \psi^*_{\vec{k}} +\Sigma_1 & \Sigma_0\eematrix
\right]^{-1}
\ee
Note that Region $I$ is also the reduced Brillouin zone as depicted in Fig(\ref{fig:k_points}) (b). 
Choice of $\psi_{\vec{k}}$ distinguishes
DCA and CDMFT. For DCA, $\psi_{\vec{k}} = \epsilon_{\vec{k}} = -2t (\cos k_x + \cos k_y)$; for
CDMFT, $\psi_{\vec{k}} = \phi_{\vec{k}} \equiv -t [1 + e^{-2 i k_x} + e^{-2 i (k_x - k_y)} + e^{-2 i (k_x + k_y)}]$ 
(see Eq(10) in Ref\cite{Fuhrmann_07}). The partial density of states in this formalism is defined as
\be
D_{S(P)}(\epsilon)  =
\frac{2}{\pi} \times \int_{\vec{k} \in I} (dk)\left[
\epsilon - i 0^+ - \bbmatrix{cc} 0 & \psi_{\vec{k}} \\ \psi^*_{\vec{k}}  & 0\eematrix
\right]^{-1}_{11(22)}
\ee
Note that the off-diagonal terms are zero for both cases.
For $\psi_{\vec{k}} = \epsilon_{\vec{k}}$ one retains the definition in DCA; for 
$\psi_{\vec{k}} = \phi_{\vec{k}}$, one obtains the PDOS in cellular DMFT as
\be
D(\epsilon)_{S(P)} = 2\times \int_{\vec{k} \in I} (dk) \,\, \delta(\epsilon -(+) Re[\phi_{\vec{k}}])
\ee


\section{Strong coupling, zero temperature limit in 2-site DCA}
In this appendix we analyze the polaron problem in the large $U$, zero temperature limit. To facilitate the
calculation, we assume that the partial density of states for $S$ and $P$ sectors to be semicircular
with the average position at negative and positive $\alpha$, i.e.
\be
D_{S(P)}(\epsilon) = \frac{1}{2\pi t^2} \sqrt{4 t^2 - (\epsilon+(-)\alpha)^2}
\ee
because of the analytical expressions.
We will show that in this limit, the self energies in $S$ and $P$ sectors indeed have pole as
\be
\Sigma_{S(P)}(\omega) = \frac{U^2}{\omega-(+)\Delta}
\ee
with the pole $\Delta = \alpha$.

We first solve the impurity cluster problem.
Following the notation in Sec.III, in the $T=0$, large $U$ limit, $(Q_1,Q_2)$  are either $ (+U,-U)$ or $ (-U,+U)$.
Averaging these two configurations, one obtains
\be
\hat{G}^{imp}=\frac{1}{a_0^2-a_1^2-U^2}
\bbmatrix{cc} a_0 & -a_1 \\ -a_1 & a_0 
\eematrix
\ee
and interpreting 
\be
G^{imp}_{S(P)} = \frac{1}{a_0 +(-) a_1 - U^2/[a_0 -(+) a_1] }
\ee
The self energy is obtained by $\hat{\Sigma} = \hat{a} - (\hat{G}^{imp})^{-1}$. After a little algebra,
one gets
\be
\Sigma_{S(P)} = \frac{U^2}{a_0 -(+) a_1 }
\ee

Using this self energies, the lattice Green's function for $S$ sector is
\be
G_S^{lat}(\omega) &=& \int \frac{D_S(\epsilon)\,\, d\epsilon}{\omega-\epsilon-\Sigma_S} \nonumber \\
&=& \frac{1}{2t^2}\left[ (\omega+\alpha-\Sigma_S) - \sqrt{(\omega+\alpha-\Sigma_S)^2-4t^2} \right]\nonumber \\
&\xrightarrow{\omega \rightarrow \infty}&  1/(\omega+\alpha-\Sigma_S) 
\ee
Similarly $G_P^{lat}(\omega) \sim 1/(\omega-\alpha-\Sigma_P) $. Comparing the lattice and local Green's function,
one gets $a_0 +(-) a_1 = \omega+(-)\alpha$ thus 
\be
\Sigma_{S(P)} = \frac{U^2}{a_0 -(+) a_1 } \sim \frac{U^2}{ \omega-(+)\alpha}
\ee



\end{document}